\documentclass[amsmath,amssymb,pra,twocolumn]{revtex4}
\usepackage[cp1251]{inputenc}
\usepackage[english]{babel}
\usepackage{color}
\usepackage{graphicx}
\usepackage{dcolumn}
\usepackage{braket}
\usepackage{amsmath}

\usepackage{natbib}

\begin{document}
\bibliographystyle{unsrt}

\title{Lorentz-invariant mass and entanglement of biphoton states}

\author{S. V. Vintskevich$^1$, D. A. Grigoriev$^{1,2}$, M. V. Fedorov$^{2,3\,*}$}
\affiliation{$^1$Moscow Institute of Physics and Technology,
Institutskii Per. 9, Dolgoprudny, Moscow Region 141700, Russia\\
$^2$A.M. Prokhorov General Physics Institute, Russian Academy of Science, Moscow, Russia\\
$^3$National Research University Higher School of Economics, 20Myasnitskaya Ulitsa,
Moscow, 101000, Russia\\
$^ * {\rm fedorovmv@gmail.com}$}

\begin{abstract}
The concept of the Lorentz-invariant mass of a group of particles is shown to be applicable to biphoton states formed in the process of spontaneous parametric down conversion. The conditions are found when the Lorentz-invariant mass is related directly with (proportional to) the Schmidt parameter $K\gg 1$ determining a high degree of entanglement of a biphoton state with respect to transverse wave vectors of emitted photons.

\end{abstract}


\maketitle

\section{Introduction}
As known \cite{Okun1,Okun2}, in the relativistic physics the mass $m$ of a group of particles is determined by the ``energy-mass-momentum$"$ interrelation
\begin{gather}
 \nonumber
 m^2c^4 = \Big(\sum_i\varepsilon_i\Big)^2 - \Big(\sum_i\vec{p}_i\Big)^2c^2
 \equiv \varepsilon_{\rm tot}^2-c^2\vec{p}_{\rm tot}^{\;2}\\
 \label{mass}
 \equiv   c^2 {\bf P}_{\rm tot}\cdot {\bf P}_{\rm tot},
\end{gather}
where $i$ numerates particles,
$\varepsilon_{\rm tot}$ and $\vec{p}_{\rm tot}$ are the total energy and momentum of the group, and
${\bf P}_{\rm tot}$ is the total 4-momentum. Clearly, defined in this way, the mass $m$ is Lorentz-invariant, as the expression in the second line of Equation ({\ref{mass}}) is proportional to the squared ``length$"$ of the 4-momentum of the group, which is invariant with respect to rotations in the 4-dimensional Minkowski  space. Moreover, as argued by L.B. Okun \cite{Okun2}, Equation (\ref{mass}) provides the only reasonable definition of a mass ``compatible with the standard language of relativity theory$"$.

\begin{figure}[h]
\includegraphics[width=8cm]{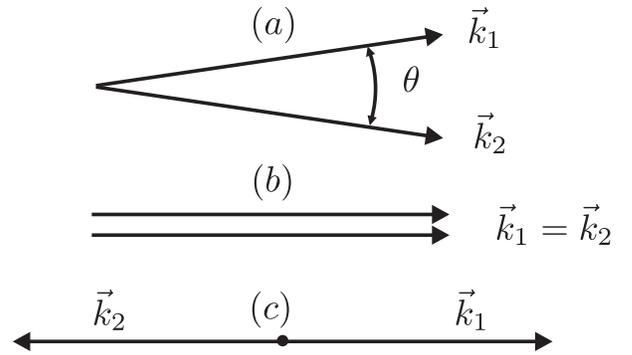}
\caption{Wave vectors of two photons in a general case $(a)$ and in the cases of collinear propagation $(b)$, and counter-directed orientation $(c)$.  } \label{fig1}
\end{figure}

The definition of Equation (\ref{mass}) is valid both for particles with  masses and for groups of massless particles, photons. For a single photon with the energy $\hbar\omega$ and momentum $\hbar k=\hbar\omega/c$, Equation (\ref{mass}) gives immediately $m_{\rm single\,phot}=0$, as it has to be. But for groups of photons the mass of a group can be different from zero if the absolute value of the vectorial sum of momenta $\hbar\sum_i\vec{k}_i$ is less than $\hbar\sum_i\omega_i/c$. The simplest example of this kind is a pair of noncollinear photons with equal frequencies $\omega$ and some angle $\theta$ between directions of their wave vectors (Figure \ref{fig1}), for which Equation \ref{mass}) yields
\begin{equation}
 \label{m(theta)}
 m(\theta)  = \frac{2\hbar\omega}{c^2}\sin\left(\frac{\theta}{2}\right).
\end{equation}
At $\theta=0$ (the case $b$) propagation of photons is collinear and the mass of the group equals zero. The mass $m(\theta)$ is maximal at $\theta=\pi$ (the case $c$) and
\begin{equation}
 \label{m-max}
 m_{\max}=m(\pi)  = \frac{2\hbar\omega}{c^2}=\frac{4\pi\hbar}{c\lambda},
\end{equation}
where $\lambda$ is the photon wavelength. Numerically, the maximal mass of two photons is rather small. E.g., at $\lambda=1 \mu$ Equation (\ref{m-max}) gives $m_{\max}\sim 4\times 10^{-33} {\rm g}$, which is about 5-6 orders of magnitude smaller than the electron mass $\sim 10^{-27} {\rm g}$. However in a system of $N$ photon pairs with the same wave vectors $\vec{k}_1$ and $\vec{k}_2$, the mass of the group becomes $N$ times larger and can become comparable with or even larger than the electron mass.

Existence of a nonzero mass means immediately that the system under consideration moves as a whole with a speed $\overline{v}$ smaller than the speed of light $c$. In a general case of any group of particles its mean velocity of motion is given by:
\begin{equation}
 \label{mean-v}
 \overline{v}=\frac{c^2 p_{\rm tot}}{\varepsilon_{\rm tot}}=c\left(1-\frac{m^2}{\varepsilon_{\rm tot}^2}\right)^{1/2}.
\end{equation}
Evidently, if $m\neq 0$ the mean propagation speed of a group of particles is smaller than the speed of light, $\overline{v}<c$. For two-photon states, decreasing of the mean velocity owing to noncollinearity of photon propagation was seen experimentally \cite{Padgett}.

For any object with $m\neq 0$ and $\overline{v}<c$ there is a frame where its mean velocity turns zero, i.e. the rest frame ($\rm r.f.$). Of course, for groups of particles in the rest frame particles can move but only in such a way that the vectorial sums of their velocities and momenta compensate each other and give zeros in the sums, i.e.
\begin{equation}
 \label{in-r-fr}
 \vec{p}_{\rm tot}^{\,\,\rm r.f.}=0,\quad \overline{v}^{\,\,\rm r.f.}=0.
\end{equation}
This clarifies the physical meaning of the Lorentz-invariant mass of a group of particles: multiplied by $c^2$, the mass $m$ coincides with the total energy of a group in its rest frame
\begin{equation}
 \label{m-in-rf}
 m=\frac{\varepsilon_{\rm tot}^{\rm r.f.}}{c^2}.
\end{equation}
An example of the rest frame for a pair of photons is that shown in Figure \ref{fig1}$c$.

As classical light fields consist of photons, the definition of the Lorentz-invariant mass is applicable also to light pulses considered as relativistic objects \cite{LP,EPL}. In this work we will consider manifolds of photons produced in nonlinear birefringent crystals under the action of a classical pump in the process of Spontaneous Parametric Down-Conversion (SPDC). As known, such photons can be entangled. The key question to be addressed below is whether there is any connection between the Lorentz-invariant mass of SPDC photons and the degree of their entanglement.

As for the pump, its Lorentz-invariant mass has to be found too, at least as the benchmark for comparison with that of SPDC photons. For the pump we can use the result of our previous work \cite{EPL}: in Equation (3) of Ref. \cite{EPL} the mean propagation speed of a diverging Gaussian pulse was found to be given by
\begin{equation}
 \label{EPL-mean-v}
 \overline{v}=c\left(1-\frac{\lambda_p^2}{8\pi^2w_p^2}\right),
\end{equation}
where $w_p$ is the pump waist. By comparing this result with that of Equation (\ref{mean-v}) at $m\ll\varepsilon_{\rm tot}$ we find the pump-pulse Lorentz-invariant mass
\begin{equation}
 \label{pump-mass}
 m_{p\,{\rm tot}}=\varepsilon_{\rm tot}\,\frac{\lambda_p}{2\pi c^2w_p}.
\end{equation}
By assuming that $\varepsilon_{\rm tot}=N\hbar\omega_p$where $N$ is the number of photons in the pump pulse, we can find from Equation (\ref{pump-mass}) the Lorentz-invariant mass per one photon
\begin{equation}
 \label{pump-mass-per 1}
 m_{p}=\,\frac{\hbar}{c\,w_p}.
\end{equation}
Qualitatively this result can be obtained from Equation (\ref{m(theta)}) with the angle $\theta$ taken equal to the diffraction-divergence angle of the pump $\theta=\lambda_p/w_p\ll 1$. As $w_p \gg \lambda_p$, the Lorentz-invariant mass of the pump per one photon is much smaller that the Lorentz-invariant mass of two counter-propagating photons (\ref{m-max}).

\section{SPDC}

Let us consider a rather simple case of the collinear frequency-degenerate regime of SPDC with the type-I phase matching. This means that the pump propagates in a nonlinear crystal as extraordinary and both emitted photons as ordinary waves. If the pump is vertically polarized, both emitted photons have the same horizontal polarization. Frequencies of the pump and of both emitted photons are assumed to be equal, correspondingly, to $\omega_p$ and $\omega_p/2$. Let $0z$ be the central pump-propagation direction, and distribution of the pump in the transverse directions ($\perp 0z$) be Gaussian with the width $w_p$: $E_p\propto \exp(-\vec{r}_{\perp p}^{\,\,2}/2w_p^2)$. The momentum-representation wave function of emitted photons depends on the transverse momenta of emitted photon $\vec{k}_{\perp 1}$ and $\vec{k}_{\perp 2}$ and is known \cite{PRA-16} to be given by
\begin{gather}
 \nonumber
 \Psi(\vec{k}_{\perp 1}, \vec{k}_{\perp 2})=
 N\exp\Bigg[-\frac{\left(\vec{k}_{\perp 1}+\vec{k}_{\perp 2}\right)^2w_p^{\,2}}{2}\Bigg]\times\\
 \label{wave func}
 {\rm sinc}\left[\frac{L\lambda_p}{8\pi n_o}\left(\vec{k}_{\perp 1}-\vec{k}_{\perp 2}\right)^2\right],
\end{gather}
where ${\rm sinc}=\frac{\sin x}{x}$, $L$ is the length of a crystal (along the $z$-axis), $n_o$ is the refractive index of the ordinary wave in the crystal, $N$ is the normalization factor.

It should be noted that, in principle, anisotropy of the refractive index of the extraordinary pump wave, $n_p$, can give rise to the additional term in the argument of the sinc-function,
$f_{\rm anisotr}(\vec{k}_{\perp 1},\vec{k}_{\perp 1})=L n_p^\prime (k_{1\,x}+k_{2\,x})/4\pi$, where $n_p^{\prime}$
is the derivative of the refractive index $n_p$ over the angle $\vartheta$ between the pump wave vector ${\bf k}_p$ and the optical axis of a crystal $OA$, with $OA$ assumed to be located in the $(x,z)$-plane. If $f_{\rm anisotr}$ is not small (compared to 1), it plays a very important role by providing anomalously high degree of angular entanglement of SPDC photons \cite{PRL-07},\cite{PRA-08}. But in this work we  assume that the term $f_{\rm anisotr}$ is small and can be ignored. Specifically, the condition justifying this assumption has the form $L|n_p^\prime|\ll w_p$, where typically $|n_p^\prime|\sim 0.1$. In the opposite case of strongly pronounced anisotropy, entanglement of SPDC photons was investigated in the works \cite{PRL-07,PRA-08}, and the arising in this case peculiarities of derivation of the Lorentz-invariant mass will be considered elsewhere separately.

The wave function (\ref{wave func}) can be used for finding mean values of the biphoton momentum
$\hbar\braket{{\vec k}_1+{\vec k}_2}$ and, finally, the Lorentz-invariant mass $m_{\rm biph}$. Because of the axial symmetry of the expression (\ref{wave func}), mean transversal components of wave vectors are equal zero, $\braket{{\vec k}_{\perp\,1}+{\vec k}_{\perp\,2}}=0$, whereas for the mean sum of longitudinal momenta in the paraxial approximation we get
\begin{gather}
 \nonumber
 \braket{k_{z\,1}+k_{z\,2}}=\\
 \nonumber
 \frac{\omega_p}{c}-
 \frac{c}{2\omega_p}\left[\braket{({\vec k}_{\perp\,1}+{\vec k}_{\perp\,2})^2}+
 \braket{({\vec k}_{\perp\,1}-{\vec k}_{\perp\,2})^2}\right]=\\
 \label{mean-mom-bph}
 \frac{\omega_p}{c}-
 \frac{c}{2\omega_p}\left[\braket{{\vec q}_+^{\,\,2}}+\braket{{\vec q}_-^{\,\,2}}\right],
\end{gather}
where $\vec{q}_\pm= {\vec k}_{\perp\,1}\pm {\vec k}_{\perp\,2}$.

In terms of two 2D vectors $\vec{q}_+$ and $\vec{q}_-$ averagings are understood as
$\braket{...}=\frac{1}{4}\int d\vec{q}_+d\vec{q}_-|\Psi|^2 (...)$, with $\frac{1}{4}$ being the transition Jacobian from variables ${\vec k}_{\perp\,1},\, {\vec k}_{\perp\,2}$ to $q_+,\,q_-$, and with the squared wave function (\ref{wave func}) taking the form
\begin{equation}
 \label{Psi-qq}
 |\Psi|^2=
 N^2\exp\left(-\vec{q}_+^{\,\,2} w_p^{\,\,2}\right) {\rm sinc}^2\left[\frac{L\lambda_p}{8\pi n_o}\vec{q}_-^{\,\,2}\right].
\end{equation}
In this expression terms depending on $\vec{q}_+$ and $\vec{q}_-$ are factorized, which is very convenient for averagings.

The first step of calculations consists in finding the normalization factor from the condition
\begin{equation}
 \label{normalization}
 \frac{1}{4}\int d\vec{q}_+\int d\vec{q}_-|\Psi|^2=1.
\end{equation}
Both integrals over $\vec{q}_+$ and $\vec{q}_-$ in (\ref{Psi-qq}) are easily calculated separately to give, correspondingly,
$\pi/w_p^2$ and $\frac{8\pi^2 n_o}{L\lambda_p}\int_0^\infty dx\,{\rm sinc}^2(x)$ with the last integral over $x$ equal to $\pi/2$. Combined together, these results give finally
\begin{equation}
 \label{Norm}
 N=\frac{w_p}{\pi^2}\sqrt{\frac{L\lambda_p}{n_o}}.
\end{equation}

The next steps are finding $\braket{{\vec q}_+^{\,\,2}}$ and $\braket{{\vec q}_-^{\,\,2}}$ in Equation (\ref{mean-mom-bph}). The first of these two quantities is determined by the integral of the exponential function in Equation (\ref{Psi-qq}) giving
\begin{equation}
 \label{q-plus-sq-av}
 \braket{{\vec q}_+^{\,\,2}}=\frac{1}{w_p^2}.
\end{equation}
As for the second term, $\braket{{\vec q}_-^{\,\,2}}$, it can be reduced in a similar way to the following integral form
\begin{equation}
 \label{q-min-sq-av}
 \braket{{\vec q}_-^{\,\,2}}=\frac{8\pi n_o}{L\lambda_p}\int_0^\infty\,x\,{\rm sinc}^2(x)\,dx ,
\end{equation}
where $x$ is the integration variable equal to the argument of the sinc-function in Equation (\ref{Psi-qq}). In principle, formally, the remaining integral
in Equation (\ref{q-min-sq-av}) diverges logarithmically at $x\rightarrow\infty$. But this divergence is related
to the used above paraxial approximation. In fact, this approximation is valid only as long as $|k_{\perp\,1,\,2}|\ll\frac{\omega_p}{2c}$, $|q_-|\ll\frac{\omega_p}{c}$, and
$x\ll x_{\rm max}=\frac{L\lambda_p}{8\pi n_o}(\frac{\omega_p}{c})^2$. For this reason, the integral in Equation (\ref{q-min-sq-av}) can be estimated as $\int_1^{x_{\rm max}}\overline{\sin^2 x}\, dx/x=\frac{1}{2}\ln (x_{\rm max})$ to give
\begin{equation}
 \label{q-min-sq-av-ln}
 \braket{{\vec q}_-^{\,\,2}}=\frac{4\pi n_o}{L\lambda_p}\ln\left(\frac{\pi L}{2n_o\lambda_p}\right).
\end{equation}
Altogether Equations (\ref{mean-mom-bph}), (\ref{q-plus-sq-av}) and  (\ref{q-min-sq-av-ln}) give the following expression for the mean momentum of biphoton pairs (multiplied by the speed of light $c$)
\begin{equation}
 \label{momentum}
 \braket{c (p_{z\,1}+p_{z\,2})}=\hbar\omega_p-c\,\delta p,
\end{equation}
where
\begin{equation}
 \label{delta-p}
 \delta p=\frac{\hbar\,c}{2\omega_p}\left[\frac{1}{w_p^2}+
 \frac{4\pi n_o}{L\lambda_p}\ln\left(\frac{\pi L}{2n_o\lambda_p}\right)\right].
\end{equation}
As always $\lambda_p\ll\{w_p, L\}$, in any case $c\delta p\ll\hbar\omega_p$ and, as $\varepsilon_{\rm biph}^{\,\,2}\equiv (\hbar\omega_p)^{\,\,2}$, the difference $\varepsilon_{\rm biph}^{\,\,2}-\braket{c (p_{z\,1}+p_{z\,2})}^{\,2}$ equals approximately $2\hbar\omega_p c\,\delta p$. In this approximation the biphoton Lorentz-invariant mass per one pair of SPDC photons takes the form
\begin{equation}
 \label{mass-final}
 m_{\rm biph}=\frac{\hbar}{c}\left[\frac{1}{w_p^2}+
 \frac{4\pi n_o}{L\lambda_p}\ln\left(\frac{\pi L}{2n_o\lambda_p}\right)\right]^{1/2}.
\end{equation}

\section{Entanglement}

Transverse momenta of SPDC photons $\vec{k}_{\perp 1}$ and $\vec{k}_{\perp 2}$ are 2D continuous variables. As known \cite{CP}, the degree of entanglement in such variables can be evaluated by the Schmidt entanglement parameter $K$ defined as the inverse trace of the squared reduced density matrix. Calculation of this parameter is rather simple and straightforward for the so called double-Gaussian wave functions, having the form of a product of two Gaussian functions, one of which depends on the sum of variables and the other one - on their difference. For non-double-Gaussian wave functions functions calculation of the parameter $K$ is a problem, and even more difficult problem is its direct experimental measurement.

Another and much easier measurable  entanglement parameter $R$ was suggested for the first time in the work \cite{PRA-04}. This parameter is  defined mathematically as the ratio of widths of the unconditional to conditional probability densities depending on the variable of one of two particles (e.g., photons). Defined in this way the parameter $R$ was found to coincide exactly with the Schmidt parameter $K$ \cite{JPB-06}, and in other cases to be on the order of $K$ \cite{Silb}. In experiment one has to split the original beam of biphotons for two channels and perform two kinds of measurements. At first, photons can by counted by a single scanning detector in only one of two channels to plot the single-particle distribution and to find its single-particle width. In the second series of measurements one has to use two detectors located in different channels, one of them scanning and another one kept at a constant position, and only coinciding signals have to be registered at the computer. In this way one gets the coincidence (or conditional) distribution and measures its width. The ratio of found widths is the parameter $R$.

It's known also that if the wave function has the form of a product of two terms, one of which depends on the sum and the other one - on the difference of variables and if the widths of these distributions are $a$ and $b$, then the coincidence and single-particle widths are, correspondingly, $\min\{a,\,b\}$ and $\max\{a,\,\}$ and, consequently, $R=\max\{a,\,b\}/\min\{a,\,b\}$. Applied to the wave function of the  form (\ref{wave func}) and variables $k_{1, x}$, $k_{2, x}$, these definitions give $a=1/w_p$, $b=2\pi\sqrt{n_o/L\lambda_p}$ and
\begin{equation}
 \label{degr-ent}
 K\sim R=\frac{\Delta k_{1, x}^{(s)}}{\Delta k_{1, x}^{(c)}}\sim
 \frac{\max\left\{1/w_{p\,},\,2\pi\sqrt{n_o/L\lambda_p}\right\}}{\min\left\{1/w_{p\,},\,2\pi\sqrt{n_o/L\lambda_p}\right\}}.
\end{equation}
This derivation and the derived expression confirm the known result \cite{PRL-07, PRA-08} that the degree of entanglement is high either if $w_p\ll\sqrt{L\lambda_p}$ or if $w_p\gg\sqrt{L\lambda_p}$.
In the second of these two limiting cases the crystal is assumed to be short compared to the diffraction length of the pump, $L\ll w_p^2/\lambda_p\equiv L_d$. Under this condition
\begin{equation}
 \label{short L}
 K\sim R_{\rm short\,L}\sim \frac{2\pi w_p\sqrt{n_o}}{\sqrt{L\lambda_p}}\gg 1.
\end{equation}
With this expression for the entanglement parameters $K$ and $R$, the derived above formula (\ref{mass-final}) for the Lorentz-invariant mass of biphoton pairs can be rewritten as
\begin{gather}
 \nonumber
 m_{\rm biph}=\frac{\hbar}{c\,w_p}\left[1+
 \frac{K^2}{\pi}\ln\left(\frac{\pi L}{2n_o\lambda_p}\right)\right]^{1/2}\\
 \approx
 \label{mass-via-K}
 \frac{\hbar K}{2 c\,w_p}\sqrt{\frac{1}{\pi}\ln\left(\frac{\pi L}{2n_o\lambda_p}\right)}\gg\frac{\hbar}{2 c\,w_p}.
\end{gather}
This result shows that in the case $L\ll w_p^2/\lambda_p\equiv L_d$ both the Lorentz-invariant mass of the biphoton state (\ref{wave func}) and its degree of entanglement are high and they are related to each other: the Lorentz-invariant mass is proportional to the Schmidt entanglement parameter $K$.

It's true, however, that this connection is not universal. In the case of very strong focusing of the pump $L\gg L_d=w_p^2/\lambda_p$, opposite to that considered above, the degree of entanglement is high too,  $K\sim\sqrt{L/L_d}\gg 1$, but the Lorentz-invariant mass (\ref{mass-final}) appears to be  independent of the degree of entanglement and determined only by the inverse waist of the pump, $m_{\rm biph}=\hbar/c\,w_p$.

\section{Conclusion}
The main result of the presented analysis concerns demonstration that there are conditions when biphotons states are highly entangled and this high entanglement is directly related to the relatively high Lorentz-invariant mass, $m_{\rm biph}\propto K\gg 1$ (\ref{mass-via-K}). We believe that this is a fundamentally important new knowledge. Though we realize that it may be difficult to imagine any ways of measuring the Lorentz-invariant mass of biphotons directly and independently of measuring coincidence and single-particle widths of momentum-distributions, finding the entanglement parameter $R\sim K$ and then using if for finding $m_{\rm biph}$ from Equation (\ref{mass-via-K}).

\section*{Acknowledgement}
The work of S.V. Vintskevich was supported by
the Foundation for the Advancement of Theoretical Physics
BASIS (PhD Student Grant No. 17-15-603-1).
D.A. Grigoriev and M.V. Fedorov acknowledge support of the Russian Foundation for Basic Researches,
grant RFBR 18-02-00634.

\bibliography{Mass-Ent-Arx}


\end{document}